# Exchange Bias Effect in Au-Fe$_3$O$_4$ Nanocomposites

Sayan Chandra[1], N.A. Frey Huls[1,2], M. H. Phan[1], S. Srinath[1,3], M. A. Garcia[4], Youngmin Lee[5], Chao Wang[5], Shouheng Sun[5], Òscar. Iglesias[6], and H. Srikanth[1,*]

[1]Physics Department, University of South Florida, Tampa, FL 33620 USA

[2]Material Measurement Laboratory, National Institute of Standards and Technology, Gaithersburg, MD 20899 USA

[3]School of Physics, University of Hyderabad, Hyderabad 500046, India

[4]Instituto de Cerámica y Vidrio - CSIC & IMDEA Nanociencia, 28049 Madrid, Spain

[5]Department of Chemistry, Brown University, Providence, RI 02912 USA

[6]Departament de Física Fonamental and Institut de Nanociència i Nanotecnologia (IN$^2$UB), Universitat de Barcelona, Av. Diagonal 647, 08028 Barcelona, Spain

ABSTRACT

We report exchange bias (EB) effect in the Au-$Fe_3O_4$ composite nanoparticle system, where one or more $Fe_3O_4$ nanoparticles are attached to an Au seed particle forming "dimer" and "cluster" morphologies, with the clusters showing much stronger EB in comparison with the dimers. The EB effect develops due to the presence of stress in the Au-$Fe_3O_4$ interface which leads to the generation of highly disordered, anisotropic surface spins in the $Fe_3O_4$ particle. The EB effect is lost with the removal of the interfacial stress. Our atomistic Monte-Carlo studies are in excellent agreement with the experimental results. These results show a new path towards tuning EB in nanostructures, namely controllably creating interfacial stress, and open up the possibility of tuning the anisotropic properties of biocompatible nanoparticles via a controllable exchange coupling mechanism.

There has been a lot of interest in the search for multifunctional nanocomposites which forecast a promising future for the next generation nanotechnology.[1,2] The goal is to artificially synthesize or fabricate a multicomponent nanostructure with controlled optical, electromagnetic and magnetic responses to suite various applications. Recently, with the advances in different chemical synthesis techniques, it has been demonstrated that one can design and chemically grow bifunctional nanostructures, called magnetic-plasmonic nanoheterostructures (MP-NHs), which are both magnetically and optically active.[1-3] Such heterostructures can be synthesized into a core/shell geometry, where the magnetic component (core) is contained within a chemically inert noble metal (shell), thereby enabling surface functionalization with biomolecules, while maintaining chemical stability against surface oxidation[4] or in different arrangements, namely heterodimers, nanoflowers, nanotriangles, nanotadpoles etc.[1,3,5,6] Interestingly, the epitaxial growth of one component on the other at the nano-scale leads to the manifestation of fascinating properties[7-9]. However, little is known about the influence of the shape and size of the constituent nanoparticles on their functional properties, knowledge of which is essential to attain better control.

In particular, Au-$Fe_xO_y$ MP-NHs have attracted a lot of attention and are thought of as potential candidates for applications in magnetic resonance imaging, magnetic/photo-induced hyperthermia, DNA sensing, cell sorting by method of magnetic separation, etc.[7,9] Recent reports have shown evidence of direct interaction due to spin polarization transfer between the magnetic moment and the non-magnetic plasmonic counterpart, thereby inducing finite magnetization in Au.[4] Simultaneously, it has been reported that the magnetic properties of $Fe_3O_4$ are influenced due to the presence of Au in direct contact, which is evidenced as exchange bias

(EB) effect, modified magnetization response to alternating fields, enhanced blocking temperature etc.[8, 10]

The EB effect in nanostructures has been an area of intense research over the last few decades.[11, 12] It manifests as an horizontal shift in the hysteresis loop accompanied by an increase in coercivity after cooling in a magnetic field, a well-known phenomenon observed typically in ferromagnet-antiferromagnet (FM-AFM) bilayer films[12]. EB has been reported also in many core-shell nanoparticles of different compositions, most notably Co-CoO[13], $CrO_2$-$Cr_2O_3$[14], FeO-$Fe_3O_4$[15], and Fe-$\gamma Fe_2O_3$[16]. However, loop shifts have also been reported in single component magnetic nanoparticles ($NiFe_2O_4$[17], $CoFe_2O_4$[18]) and they are usually attributed to a "shell" of disordered surface spins formed in nanoparticles with a high surface-to-volume ratio that results in EB due to coupling to core spins. The origin of this behavior is the fraction of surface spins with decreased co-ordination (and thus weaker bonding) that increases when particle size decreases. These disordered spins can take on a number of configurations that are quasi-degenerated in energy due to randomness and frustration induced by the competition between exchange coupling and surface anisotropy. This degeneracy can be usually broken by applying a magnetic field while cooling the particle and, thus, EB is induced. Consequently, EB can be introduced in nanostructures mainly by (i) the growth of two different magnetic phases with competing anisotropies, or (ii) diminishing the particle size of a single magnetic material (< 5 nm in case of ferrites), such that the surface anisotropy exceeds the anisotropy of the bulk material by several orders of magnitude. For example, in the case of $Fe_3O_4$, it has been theoretically predicted that EB can develop in particles with diameter ~ 2.5 nm or below, and ceases to exist for bigger particle sizes.[19]

In this article, we show a remarkable EB effect in Au-Fe$_3$O$_4$ composite nanoparticles, where Fe$_3$O$_4$ nanoparticles were epitaxially grown on one or more facets of the Au particle forming the corresponding "dimer" and "cluster" morphologies. The observation of exchange bias in case of 8 nm Au – 9 nm Fe$_3$O$_4$ nanostructures is rather perplexing. For Fe$_3$O$_4$ nanoparticles with mean size ~ 8 nm, it has been reported that no additional surface or shape anisotropy exists and, hence, no EB effect is observed.[20] This suggests that, in the Au-Fe$_3$O$_4$ system, EB is induced in ~ 9 nm Fe$_3$O$_4$ by the epitaxial growth of a non-magnetic metal on it. This phenomenon of the development of EB in magnetic nanoparticles, just by the introduction of a non-magnetic material on its surface is uncommon and its origin from a fundamental point of view is not understood. We investigate this by performing a series of systematic magnetometry experiments along with atomistic Monte-Carlo simulations to gain insights into the complex physics associated with this system leading to EB.

**RESULTS AND DISCUSSION**

8 nm Au – 9 nm Fe$_3$O$_4$ dimer- and cluster-shaped nanoparticles were synthesized following the method outlined in reference.[5] For comparison purposes, 9 nm spherical Fe$_3$O$_4$ nanoparticles of the same size were synthesized following the same procedure in the absence of Au seed particles. We also examined Fe$_3$O$_4$ particles that were obtained after the chemical etching of the Au off the Au-Fe$_3$O$_4$ dimers resulting in a dented morphology. Figures 1a and 1b show the dimer and cluster nanoparticles respectively, while Figure 1c shows the Fe$_3$O$_4$ particles after the chemical etching of the Au from the dimers. High resolution transmission electron microscopy of a dimer particle (inset of Figure 1a) reveals the epitaxial growth of single crystalline Fe$_3$O$_4$ on the Au seed. It has been reported that Au-Fe$_3$O$_4$ composite nanoparticles show a red-shift of the surface plasmon of the Au owing to an electron deficient population arising from its contact with the $Fe_3O_4$.[5] We avail this optical property as a tool to characterize the quality of the samples. Consistent with previous reports, the surface plasmon resonance (SPR) absorption spectra for the dimers (Figure 1d) exhibits a red shift of the absorption peak compared to that of the Au seeds which have an absorption frequency of ~ 520 nm. For the Au-$Fe_3O_4$ clusters, it is worth noting that no meaningful results could be obtained from SPR measurements, the electronic oscillations being so damped by the modified electronic structure as to render the quantitative results unusable. This indicates enhanced interface communication between Au and $Fe_3O_4$.[9]

Figure 2 shows the temperature dependence of magnetization in an applied field of 0.01 mT after zero-field cooled (ZFC) and field-cooled (FC) procedures. In the limiting case of single-domain monodispersed particles, the ZFC magnetization exhibits a peak ($T_B$) which is associated with the blocking phenomenon of the nanoparticles. The spherical particles have $T_B$ ~ 36 K, and, although, the dimers and the clusters are constituted by similar ~ 9nm $Fe_3O_4$ particles, they exhibit an increase in $T_B$ up to ~ 65 K (Fig. 2b) and 90 K (Fig. 2c) respectively. This feature, together with the increase in $T_B$ also associated with broadening of the peak, are typically associated with enhancement of magnetic anisotropy or inter-particle interactions.[21, 22] The contributions of the shape and surface anisotropies to the effective magnetic anisotropy play a vital role in determining the blocking temperature. In case of spherical $Fe_3O_4$ nanoparticles, it has been reported that the contribution of shape anisotropy nullifies owing to its symmetry.[23] As it can be seen from the TEM images in figure 1, the shape of the $Fe_3O_4$ nanoparticles in case of the dimers and the clusters deviate from that of an ideal sphere. Moreover it is not clear if the epitaxial growth of the $Fe_3O_4$ nanoparticles on the Au seeds in case of the dimers and clusters modify and alter their surface anisotropy.

The blue squares in Figure 3 show the ZFC magnetization versus magnetic field (M-H) curves for the $Fe_3O_4$ spherical particles (3a), the Au-$Fe_3O_4$ dimers (3b) and the Au-$Fe_3O_4$ clusters (3c) taken at 2 K. In the insets we provide complete FC hysteresis loops for the respective samples. Each type of particle morphology has different properties, with the M-H curve for the spheres being similar to other reports of $Fe_3O_4$ in the same size range.[24] The coercivity ($H_C$) increases with the introduction of the Au with the dimers having a higher $H_C$ (64 mT) than the spheres (39 mT) and the clusters having the highest $H_C$ of the three (124 mT). We also note for the clusters a slow approach to saturation and a low remanent magnetization ($M_R$).

Hysteresis loops were also measured for all three samples after field cooling (FC) to 2 K in 1 T and the results appear as red circles in Figure 3. While no EB is observed for the $Fe_3O_4$ nanoparticles, both the dimers and the clusters are found to exhibit substantial exchange fields ($H_E$) which were calculated using the formula:

$$H_E = \frac{-(H_{C1} + H_{C2})}{2}$$

where $H_{C1}$ and $H_{C2}$ are the coercive fields measured along the descending and ascending branches respectively. The dimers display an $H_E$ of 26 mT, while the clusters have an $H_E$ of 50 mT. The dimers and clusters also show a slight increase in coercivity as well (73 mT versus 64 mT for the dimers and 130 mT versus 124 mT for the clusters) as well as a vertical shift and increase in remanent magnetization. This phenomenon is frequently seen in exchange coupled nanoparticle systems where the pinning layer is a metastable disordered state such as a shell of disordered spins.[11, 25] While in FM-AFM systems it is the result of uncompensated interfacial spins, similarly in FM-disordered systems the net preferred direction of frozen spins lies along the cooling field creating a vertical asymmetry.[26] The EB results suggest that there is a pinning layer present somewhere in the Au-Fe$_3$O$_4$ particles (which is not present in the spherical Fe$_3$O$_4$ nanoparticles) and that this pinning is much stronger in the clusters than in the dimers.

It is well established that EB may develop in nanoparticles by way of exchange coupling of the core moments with the surface spins.[27]. A prerequisite for EB is the presence of two magnetic phases, one that reverses with the external field, and one that does not. In case of nanoparticles, if the surface anisotropy is high compared to the anisotropy of the core moments, then the surface spins may behave as pinning centers necessary for EB. In case of the dimers and the clusters, if the development of EB is indeed due to the enhancement of the surface anisotropy, then we should observe an increase in the effective magnetic anisotropy of the dimers and clusters in comparison to the spherical particles.

To elucidate this, we have performed radio-frequency transverse susceptibility (TS) experiments on these systems. The TS has been a largely successful technique for directly probing magnetic anisotropy in assemblies of single domain particles.[28, 29] This method measures the susceptibility by means of a perturbing alternating current field ($H_{AC}$) oriented perpendicular to a changing direct current (DC) magnetic field, $H_{DC}$. In the theoretical paper on the transverse susceptibility of a Stoner Wohlfarth particle, Aharoni *et al.* calculated that three singularities would be observed in the transverse susceptibility when $H_{DC}$ was scanned from positive to negative saturation.[30] Two of these peaks were located precisely at the positive and negative effective anisotropy fields ($H_K$) and the third peak at the switching field ($H_S$). In the experimental set up of TS, the sample is inserted in a coil whose axis is perpendicular to $H_{DC}$ and produces a small perturbing AC magnetic field ($H_{AC} \approx 1$ mT) perpendicular to $H_{DC}$. The coil is part of a self-resonant circuit driven by a tunnel diode oscillator which oscillates at a frequency of around 2.6 MHz. The coil is inserted into the PPMS such that it can operate within a temperature range of 5 K – 300 K and a DC magnetic field range of ±7 T. As $H_{DC}$ is swept, the permeability of the sample sitting in the coil changes and thereby changes the inductance of the coil. This results in a shift in the oscillating frequency which is recorded within an accuracy of one 1 Hz in 1 MHz. Because the change in frequency of the circuit is a direct consequence of the change in inductance as the sample is magnetized, $\Delta f$ is directly proportional to $\Delta \chi_T$. We therefore most interested in the quantity

$$\frac{\Delta \chi_T}{\chi_T}(\%) = \frac{|\chi_T(H) - \chi_T^{sat}|}{\chi_T^{sat}} \times 100$$

as a function of $H_{DC}$ where $\chi_T^{sat}$ is the TS at the saturating field $H_{sat}$. This quantity, which represents a figure of merit, does not depend on geometrical parameters and is useful for comparing the TS data for different samples, or for the same sample under different conditions.

The left panel of Figure 4 shows the magnified view of the bipolar TS curves (positive saturation field to negative saturation field and back) of the spheres (a), dimers (b) and clusters (c) to illustrate the peaks corresponding to the anisotropy fields (±$H_K$). In the middle panel (Figure 4 d, e, f), a family of unipolar TS scans at selected temperatures are provided for the three samples. We observe that while the anisotropy peaks for the spherical particles are symmetric, the dimers exhibit asymmetric TS curves, and the asymmetry increases in case of the clusters. In experimental measurements of arrays of nanoparticles, certain deviations are observed such as asymmetry and broadening of the ±$H_K$ peaks and the merging of $H_S$ with the second anisotropy peak.[13, 22] This merging, along with asymmetry in peak height and field placement require us to distinguish between the anisotropy peak observed upon reducing the field from saturation (henceforth ±$H_{K1}$) and the peak observed upon increasing the field after crossing through H = 0 (±$H_{K2}$) which is frequently the combination of the switching and anisotropy peaks. These deviations can be attributed to the size distribution in the nanoparticles and the dipolar interactions between the nanoparticles.[28] However, since $H_{K1} \neq H_{K2}$ due to the merging of the switching peak with the $H_{K2}$, we exclusively use the $\pm H_{K1}$ peak to quantify the anisotropy field, whereas we use the $\pm H_{K2}$ peak to make qualitative observations regarding the switching behavior of the nanoparticle assembly. The right panel in figure 4 shows the evolution of $\pm H_{K1}$ with temperature for the three samples. The anisotropy field value for the spherical particles (Figure 4g) at 10 K (~ 64 mT) is found to be less than that of the dimers ~ 78 mT (Figure 4h). Considering that the magnetic volume of the spherical particles and the dimers is the same, the increase in anisotropy field in case of the dimers can be attributed to enhanced surface anisotropy. This suggests that in case of the dimers, perhaps, the seed mediated growth of $Fe_3O_4$ on Au results in a different magnetic configuration of the surface spins with higher anisotropy as compared to spherical $Fe_3O_4$. Interestingly, the anisotropy field for the clusters at 10 K (~ 400 mT) is significantly higher than the dimers, which can be attributed to the complex morphology and shape of each cluster particle. Similar to previous studies, as the temperature is increased, the anisotropy field decreases for all three samples and vanishes above their respective blocking temperatures.[22, 31] Hence, the TS experiments allow us to conclusively infer that the $Fe_3O_4$ particles in the dimers and the clusters have higher surface anisotropy compared to spherical $Fe_3O_4$. To this point, *the question that arises is: how does the seed mediated growth of $Fe_3O_4$ on Au affect its surface magnetization?*

In an earlier study, we have reported the development of stress at the heterogeneous interface in Au-$Fe_3O_4$ dimers.[32] The mechanical modeling analysis revealed that the development of stress occurred due to different thermal expansion coefficients of Au and $Fe_3O_4$ at the interface of Au-$Fe_3O_4$, and was of the order of 1 – 5 GPa. It has also been reported that compacting $Fe_3O_4$ nanoparticles under external pressure (1 – 5 GPa) can result in the development of surface spin disorder and hence replicate a core/shell magnetic structure.[33] In such a scenario, EB has been observed in $Fe_3O_4$ nanoparticles up to 20 nm. We would like to point out that the magnitude of external stress applied to the $Fe_3O_4$ nanoparticles of reference 33 is of the same order as that generated across the Au-$Fe_3O_4$ interface of the dimers. We hypothesize that as a consequence of the interfacial stress, the $Fe_3O_4$ particles in the dimers develop surface spin disorder by way of energy minimization. The disordered surface spins are highly anisotropic, which is consistent with the rise in effective magnetic anisotropy of the dimers as seen in Figure 4. The disordered surface spins undergo exchange coupling with the core moments resulting in the EB effect in the dimers and the clusters. It is to be mentioned that the manifestation of the EB effect in both intrinsically and externally strained systems is not uncommon. It has been shown in case of $Pt_3Fe$ single crystals that plastic deformations can lead to superlattice dislocations, thereby inducing growth of FM domains within an AFM matrix, which couple magnetically to exhibit EB effect.[34] In another study, the EB effect was found to diminish with the application of external mechanical strain on FM/AFM heterostructures.[35]

With regard to the present study, one can argue, that if EB in the dimers is solely due to the development of interfacial stress induced surface spin disorder, then, EB should vanish if the source of the interfacial stress were to be removed. To test this, we removed the interfacial stress by chemically etching away the Au seed in the dimers, resulting in dented $Fe_3O_4$ etched particles (Figure 1c). Figure 5a shows a magnified view of the ZFC and 1 T FC hysteresis loops for the etched dimers taken at 2 K. The inset shows the complete 1T FC loop. We find that $H_C$ of the etched dimers for the ZFC loop (35 mT) is similar to the spherical $Fe_3O_4$ particles shown in Figure 3a (39 mT). In the case of the etched dimers, field cooling actually reduced the $H_C$ to 28 mT and resulted in an improvement in the remanent magnetization. We tested this result using several different cooling fields and no EB was seen in fields as low as 0.1 T. Figure 5b shows normalized ZFC hysteresis loops for the spherical, dimers and etched particles. Usually magnetic nanoparticles with negligible surface anisotropy or surface spin canting exhibit a (i) low coercive field, and (ii) quick approach to saturation. In case of the spheres and the etched dimers, we observe both properties, however, the hysteresis loop for the dimers is found to exhibit a rather slow approach to saturation. The non-saturating behavior of the dimers is consistent with the idea of highly anisotropic disordered surface spins which resist aligning even at large magnetic fields.[36] After etching, as the interfacial stress in the dimers is lost, the surface spin disorder diminishes. Hence, the etched dimers attain saturation, similar to spherical $Fe_3O_4$. It is to be mentioned that the shape asymmetry in the etched dimers does not enhance disordering of spins in the surface layer and lead to the consequent development of EB. This is in agreement with earlier experimental results proving that surface or shape contributions to effective magnetic anisotropy are negligible in spherical $Fe_3O_4$ nanoparticles down to 8 nm, giving a natural explanation for the absence of EB in this system.[20] The cartoon shown in Figure 5c summarizes the different scenarios encountered in the Au-$Fe_3O_4$ composite system. The presence of significantly disordered surface in the dimers and clusters is depicted by a darker shade.

To complement our experimental results, we have performed Monte Carlo (MC) simulations of an atomistic model of $Fe_3O_4$ for Heisenberg spins similar to our studied samples (see the Methods section for details of the simulation). The simulated hysteresis loops after a FC process at $h_{FC}$= 100 K for the spherical-, cluster-, and dimer- shaped nanoparticles are shown in the main panels of Fig. 6. In our simulations, we can recreate the effect of interfacial stress and consequential disorder in the surface spins by assigning Néel surface anisotropy with an increased value ($k_S$ = 30) as compared to the core spins which are assumed to have the same anisotropy as the bulk ($k_C$= 0.01).

The inclusion of increased surface anisotropy ($k_S$= 30, red circles in Figure 6 a, b) results in a slower approach to saturation and high field irreversibility, along with the expected observation of EB, which is practically absent for the case in which surface and core anisotropies are the same, $k_S$ = $k_C$ (blue squares in Figure 6 a,b). Moreover, the horizontal shift of the loops is noticeably higher for the clusters than for the dimers, as also observed experimentally, which demonstrates that the EB can be tuned by the increase of the contact interfaces between Au and the magnetic NPs.

The case for the etched dimers can be recreated by assigning the same anisotropy values to the surface and core moments. We observe that for $Fe_3O_4$ in the cluster and etched dimer geometries, with surface anisotropy equal to the core value ($k_S$= $k_C$= 0.01), neither loops exhibit horizontal shifts after a FC (blue squares in Figure 6 a, b). The absence of EB in the asymmetric etched dimers (dashed lines in Figure 6b) is consistent with the experimental data shown in Figure 5a.

Note that the hysteresis loops for $k_S$= 0.01 have more squared shape than those for $K_S$= 30, which display high field linear susceptibility, irreversibility and lack of saturation both for dimers and clusters, as also observed experimentally. This last point can be ascribed to the contribution of surface spins (green squares in the inset of Figure 6b) that presents an hysteresis loop typical of a frustrated material and dominates the magnetization reversal of the whole particle. However, core spins (yellow circles in the inset of Figure 6b), reverse in a more coherent fashion although influenced also by the ones at the surface. The higher value of the remanent magnetization for the dimers than for the clusters is also in agreement with the experimental results. The high degree of disorder at the particle surface is corroborated by the snapshots of the spin configurations displayed in Figure 6(c-f), where one can notice that, even after the high FC process (Figures 6c and 6e) only the core spins (drawn in lighter colors) are aligned along the anisotropy axis while surface spins remain highly disordered even at low temperatures. Snapshots taken near the coercive field (Figures 6d and 6f) demonstrate a more coherent reversal of core spins that are dragged by the disordered shell of surface spins (drawn in darker tones).

**CONCLUSIONS**

We have performed a systematic study to understand the origin of EB in 8 nm Au – 9 nm $Fe_3O_4$ nanocomposites with two configurations, namely dimers and clusters. The dimers and the clusters exhibit enhanced magnetic anisotropy compared to 9 nm spherical particles, which, has been directly probed by transverse susceptibility measurements. Exchange bias effect is observed in the dimers and the clusters, as opposed to the spherical, and the etched dimer particles. The increase in effective magnetic anisotropy and the development of EB in the dimers and clusters are attributed to the presence of highly disordered surface spins which foster as a result of stress (order of ~ few GPa) across the Au-$Fe_3O_4$ interface. We also show that EB vanishes with the removal of the interfacial stress in case of the etched dimers. Our experimental results are well supported by atomistic Monte-Carlo simulation studies which provide conclusive evidence of the manifestation of EB solely due to highly anisotropic disordered surface spins.

Our study reveals a new path to deliberately engineer EB into nanoparticles with high magnetization, by actuating a local stress in its environment in the form of a noble metal nanoparticle. Consequently, the interfacial stress, and hence the EB field can be tuned by varying the size of both the Au and $Fe_3O_4$ particles. The capability to induce controlled EB effect in $Fe_3O_4$ up to 20 nm large particles opens the possibility of various applications involving EB which, were otherwise limited by low magnetization of smaller nanoparticles. Furthermore, depending on the size of the magnetic particle in such nanocomposites, one can gain control over the onset temperature of EB as well.

**MATERIALS AND METHODS**

The Au-Fe$_3$O$_4$ dimer nanoparticles were prepared by decomposing iron pentacarbonyl, Fe(CO)$_5$, over the surface of Au nanoparticles, followed by oxidation in air. The Au nanoparticles were formed in situ by injecting HAuCl$_4$ solution into the reaction mixture. By changing the solvent from a nonpolar hydrocarbon to a slightly polarized solvent (e.g. diphenyl ether), flower-like cluster nanoparticles were synthesized. Further details of the synthesis are presented elsewhere[5]. The important feature of this synthesis is that the sizes of the Au and the Fe$_3$O$_4$ components can be independently tuned to create a variety of size combinations with the Au phase stable up to 8 nm and the Fe$_3$O$_4$ stable up to 20 nm. Samples for TEM analyses were prepared by evaporating a drop of diluted colloidal dispersion onto carbon coated copper grids. Low-resolution and high resolution TEM images were acquired respectively on a Philips EM 420 (120 kV) and a JEOL 2010 (200 kV).

Samples were prepared for magnetic measurements by evaporating nanoparticle suspensions into a gelatin capsule. Since the surfactant on the particles also acts as an adhesive, the dried particles made a paste in the gelcap and we are confident that there was no movement during measurement. Magnetization versus temperature measurements were performed on a physical properties measurement system (PPMS) by Quantum Design by first cooling the samples from room temperature in zero field to 2 K. A field of 1 mT was then applied and the magnetization measured upon warming at 2 K/min to room temperature wherein the temperature was lowered at 2 K/min to 2 K.

Field cooled M-H curves were performed on the spherical $Fe_3O_4$, Au-$Fe_3O_4$ dimers and Au-$Fe_3O_4$ clusters by cooling in a field from above room temperature to the lowest temperature and then and then incrementally increasing the temperature and carrying out a hysteresis loop measurement at each step using a PPMS. The same protocol was used for a second Au-$Fe_3O_4$ dimer sample and the etched Au-$Fe_3O_4$ dimers using a superconducting quantum interference device, magnetic properties measurement system (MPMS, Quantum Desgin).

Optical absorption measurements were performed at room temperature with a Shimadzu 3100 double-beam spectrophotometer attached with an integrating sphere in the transmission mode. Samples were deposited onto a glass substrate and placed in the beam path. The beam spot was 4 mm x 4 mm at the sample surface. A linear background was subtracted to account for other contributions to the absorption spectrum.

Monte Carlo simulations using the standard Metropolis algorithm are based on Heisenberg classical spins representing Fe ions placed on the nodes of the real crystal lattice of magnetite with the following interaction Hamiltonian:

$$H/k_B = -\sum_{\langle i,j \rangle} J_{ij} \left( \vec{S}_i \cdot \vec{S}_j \right) - \sum_i \vec{h} \cdot \vec{S}_i + E_{anis}$$

which includes the exchange interactions ($J_{ij}$), the Zeeman energy with h= $\mu H/k_B$ (H is the magnetic field and $\mu$ the magnetic moment of the magnetic ion), and the magnetocrystalline anisotropy energy $E_{anis}$. Values for the $J_{ij}$ between spins with tetrahedral and octahedral coordination have been taken from the available literature.[19, 37]

Fe ions with reduced coordination with respect to bulk are considered to be surface spins with Neél type anisotropy and anisotropy constant $K_S$, while core spins have uniaxial anisotropy along the field direction with anisotropy constant $K_C$. Therefore, $E_{anis}$ has the form:

$$E_{anis} = -k_S \sum_{i \in S} \sum_{j \in nn} \left(\vec{S}_i \cdot \hat{r}_{ij}\right)^2 - k_C \sum_{i \in C} \left(\vec{S}_i \cdot n_i\right)^2,$$

where $\hat{r}_{ij}$ is a unit vector joining spin *i* with its nearest neighbors *j* and $n_i$ is the anisotropy axis of each crystallite. The value of the anisotropy constants, expressed in units of K/spin have been taken as $K_C$ = 0.01 and $K_S$ varying in the range 0.01- 30. As for the particle geometries, the dimer particle has been modeled has a sphere of radius 5.5a ('a' is the unit cell size) truncated by a sharp facet where the magnetite contacts the Au. The cluster-shaped particle is formed by four overlapping spheres of radius 5a surrounding a spherical hole that stands for the central Au component.

FIGURES

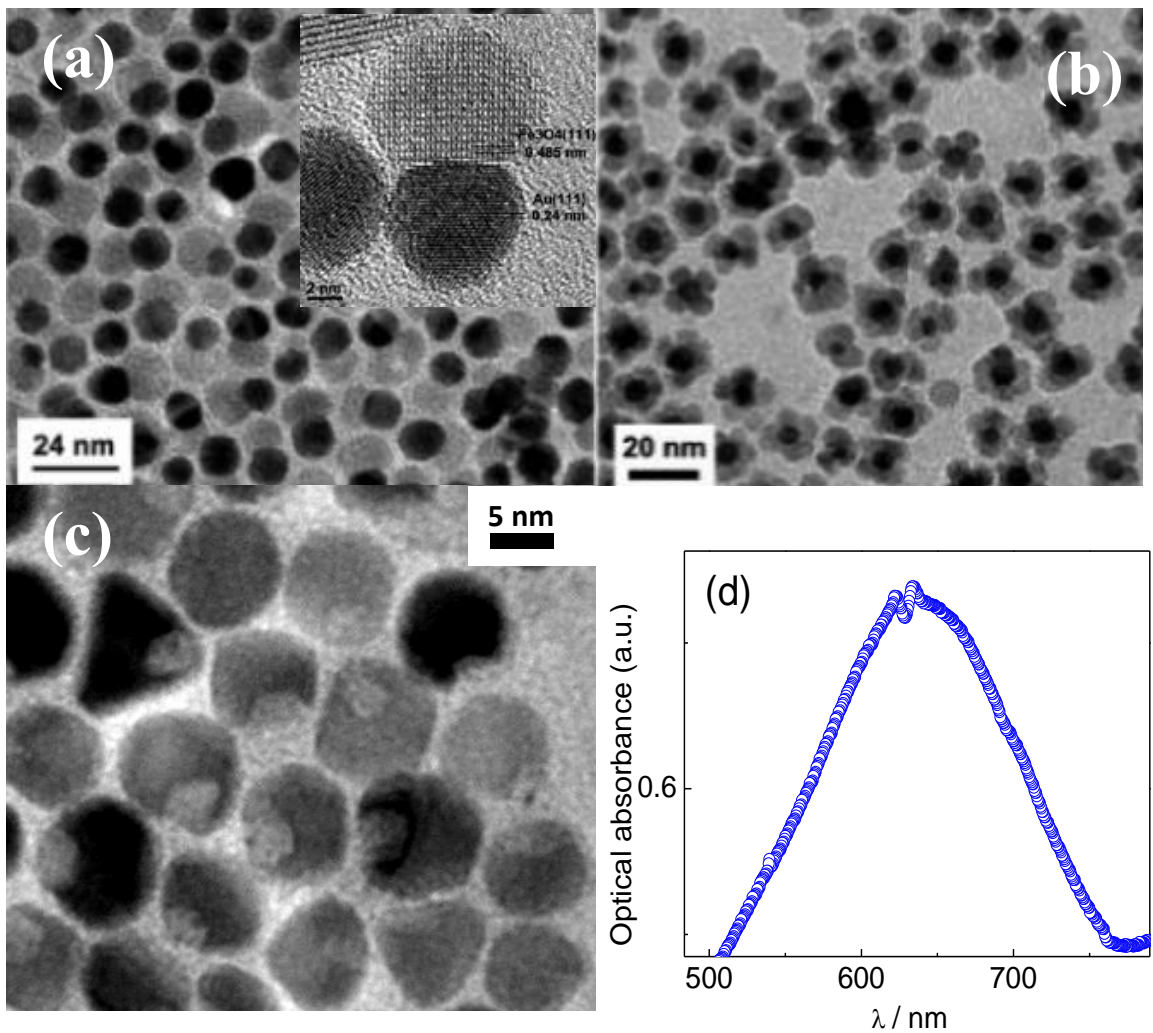

**Figure 1.** TEM images of (a) Au-Fe$_3$O$_4$ dimers, (b) Au-Fe$_3$O$_4$ clusters, (c) dimers after Au etching, and (d) Surface Plasmon resonance absorption spectra of the Au-Fe$_3$O$_4$ dimer nanoparticles. Inset of (a) shows the HRTEM image of an Au-Fe$_3$O$_4$ dimer showing single crystalline Au and Fe$_3$O$_4$ grown epitaxially.

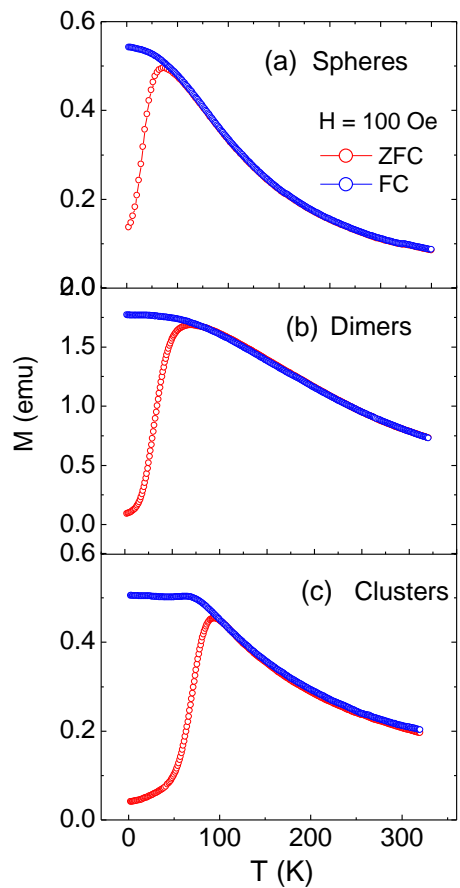

**Figure 2.** Temperature dependence of magnetization in the zero-field cooled and field cooled protocols for (a) spheres, (b) dimers, and (c) clusters.

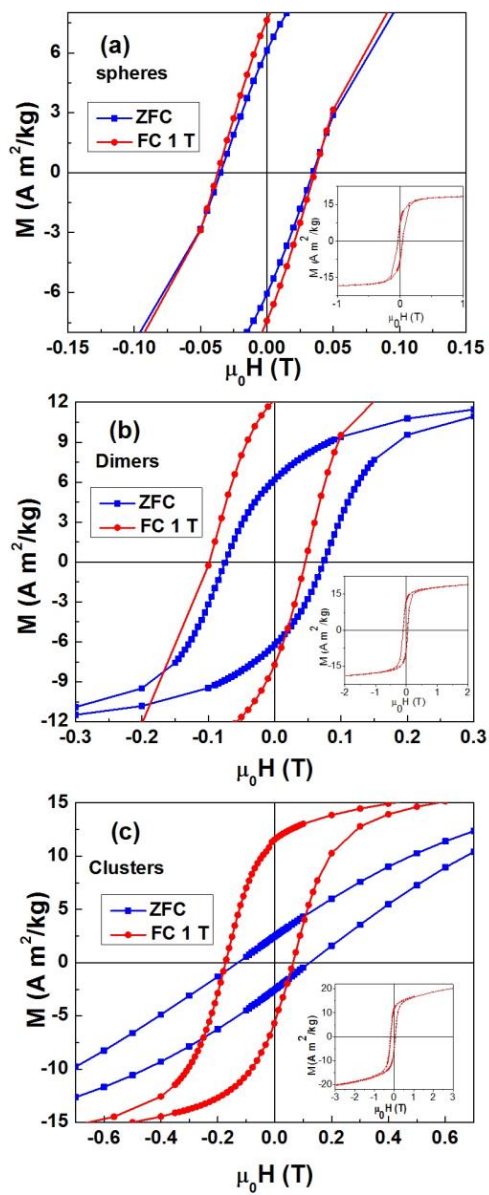

**Figure 3.** Zero field cooled (blue squares) and 1 T field cooled (red circles) magnetization versus field (M-H) curves taken at 2 K for (a) $Fe_3O_4$ spheres, (b) $Au-Fe_3O_4$ dimers, and (c) $Au-Fe_3O_4$ clusters. Insets of (a, b, c) show complete FC hysteresis loops for respective samples.

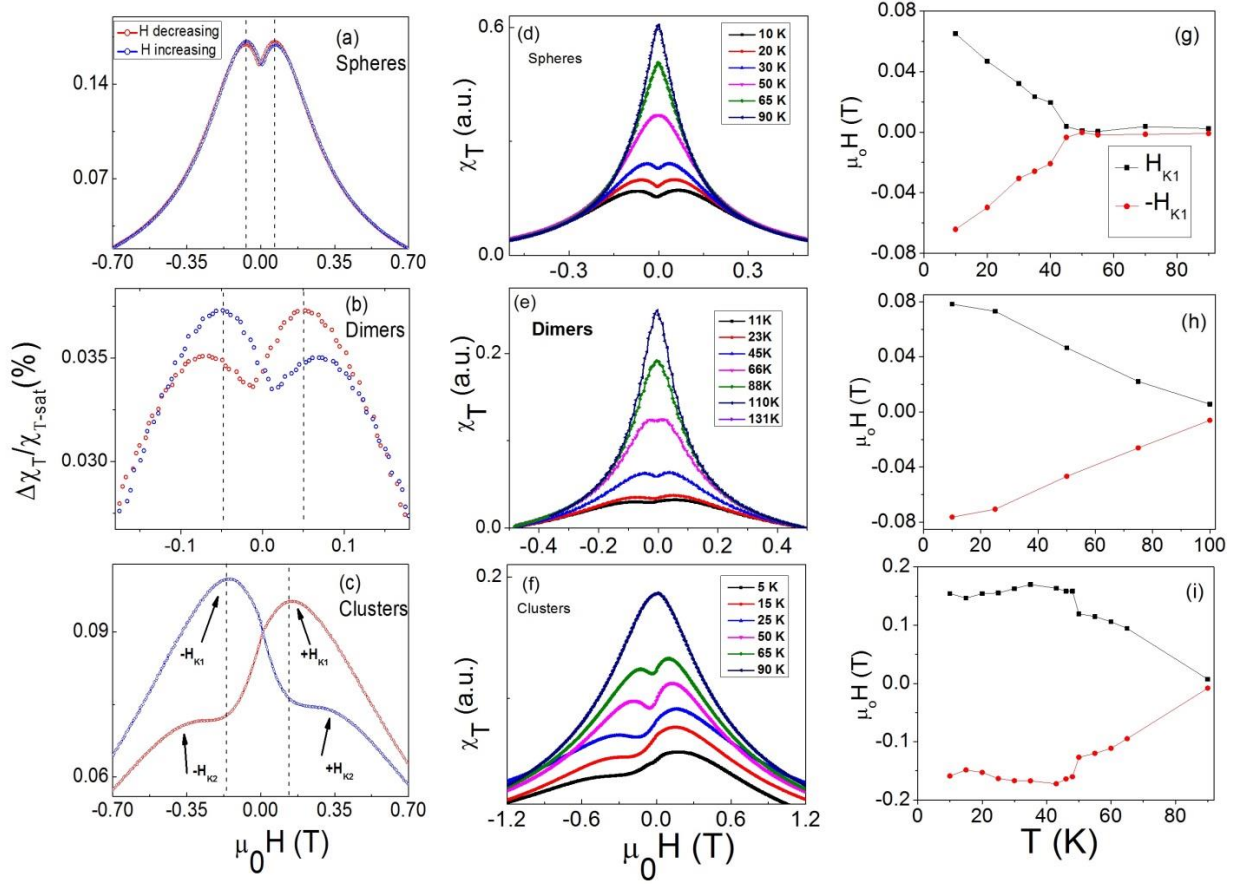

**Figure 4.** Representative bipolar TS curves measured at 20 K for (a) spheres, (b) dimers, and (c) clusters. Selected unipolar TS curves to illustrate the evolution of anisotropy peaks with temperature for (d) spheres, (e) dimers, and (f) clusters; Evolution of first anisotropy field with temperature for (g) spheres, (h) dimers and (i) clusters.

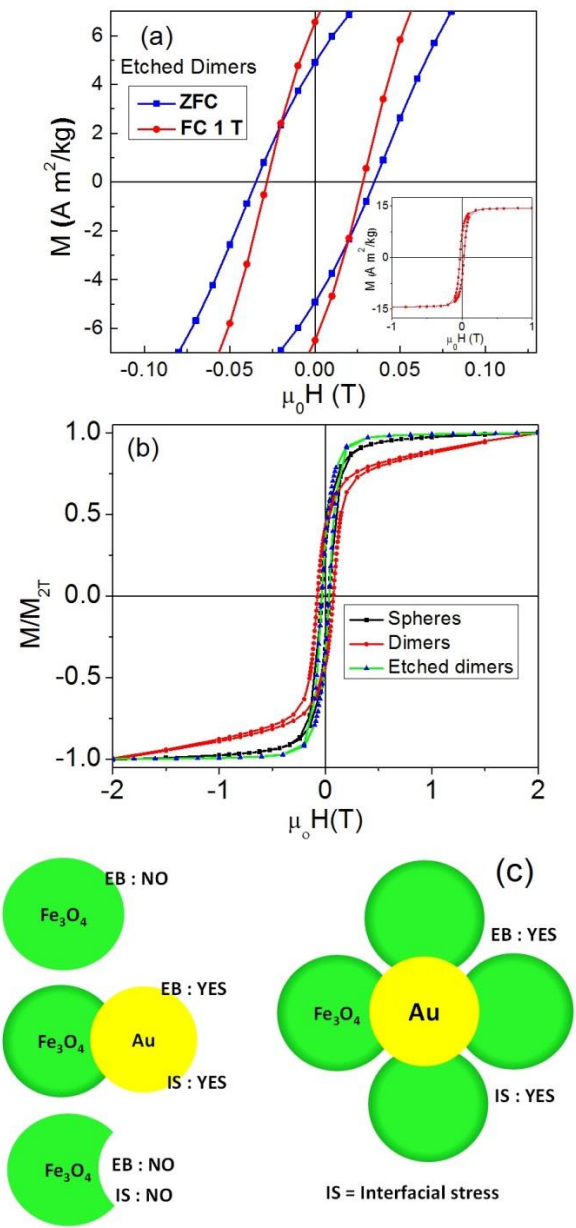

**Figure 5.** (a) Zero field cooled (blue squares) and 1 T field cooled (red circles) magnetization versus field (M-H) curves taken at 2 K for etched dimers; (b) comparison of zero-field cooled hysteresis curves for spheres, dimers, and etched dimers; (c) Schematic representation of the Au-$Fe_3O_4$ morphologies studied

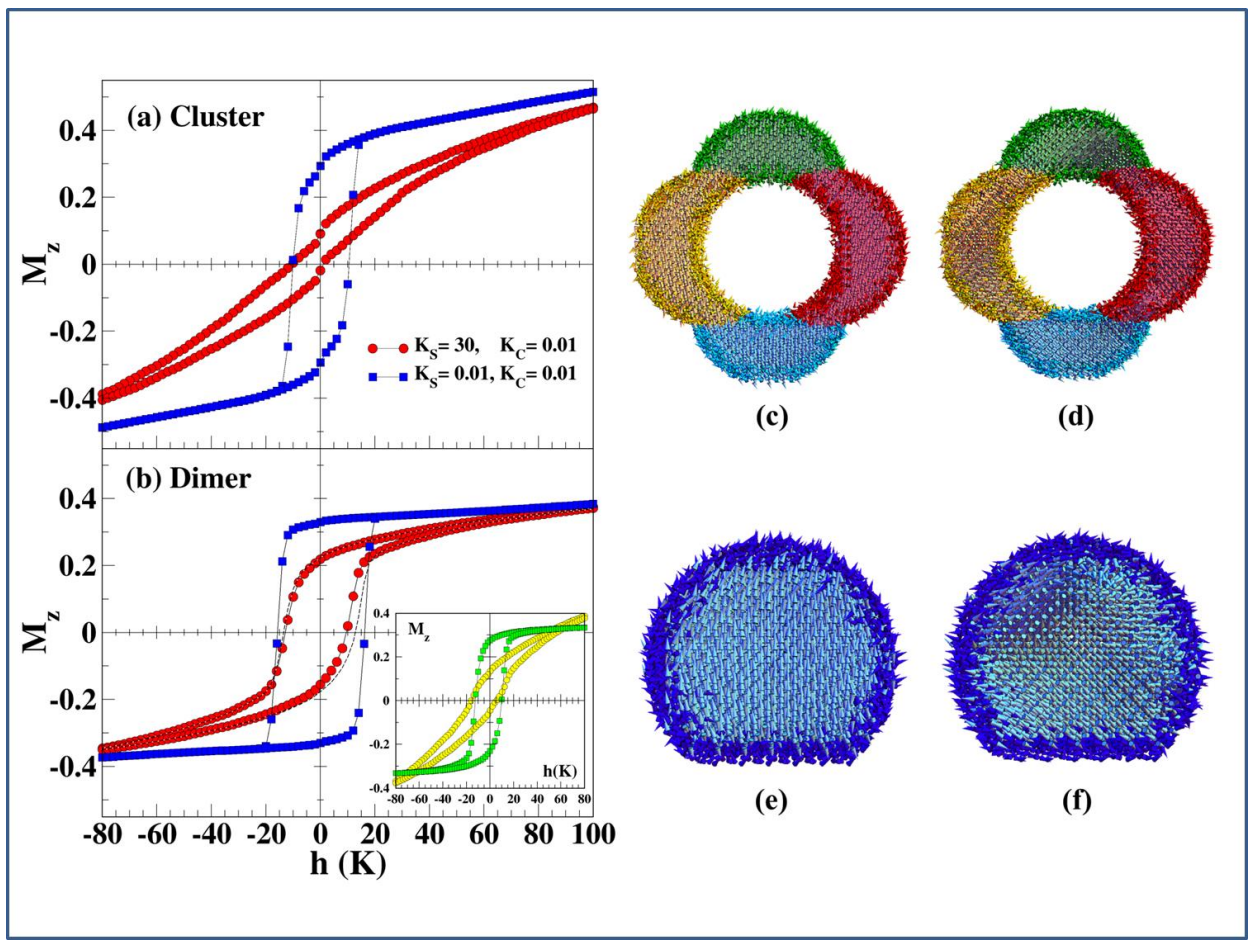

**Figure 6.** Low temperature hysteresis loops simulated after a cooling in a magnetic field $h_{FC}$ = 100 K as computed by MC simulations of individual nanoparticles with cluster (a) and dimer (b) geometries. The non-magnetic metal is simulated as a hole in the middle for the cluster geometry and a sharp facet for the cluster. Panels (a) and (b) show hysteresis loops of a particle with cluster and dimer geometry, respectively, for two different values of the surface anisotropy constant: $k_S$= 0.01 (blue squares) equal to the core value $k_C$= 0.01, and increased surface anisotropy $k_S$= 30 (red circles). The dashed lines in (b) stand for a spherical particle of the same size as the dimer. The Inset displays the contribution of the surface (yellow circles) and core (green squares) spins of a dimer particle to the hysteresis loop for $k_s$ = 30. Snapshots of the spin configurations for cluster (c and d panels) and dimer (e and f panels) particles for $k_S$= 30 (red

circles) obtained at the end of the FC process (c and e panels) and at the coercive field point of the decreasing field branch (d and f panels) of the hysteresis loops displayed in Figure 6 (a) and (b). For clarity, only a slice of width 4a along the applied field direction and through the central plane of the particles is shown. Surface spins have darker colors and core spins have been colored lighter.

ASSOCIATED CONTENT

**Supporting Information**. : TO BE FINALIZED

AUTHOR INFORMATION

**Corresponding Author**

*sharihar@usf.edu

**Author Contributions**

N.F.H., S. Srinath, H.S., and S. Sun jointly conceived the study. C. W., Y.L. and N.F.H. synthesized the nanoparticles. S.C., N.F.H., S. Srinath, and M.H.P. performed magnetic measurements and data analysis. M.A.G. performed optical measurements. Ò.I. performed magnetic simulations. S.C., N.F.H., M.H.P., Ò. I., and H.S. wrote the paper.

ACKNOWLEDGMENT

Work at USF supported by DOE through grant number DE-FG02-07ER46438. HS also acknowledges support from the Center for Integrated Functional Materials through grant USAMRMC-07355004. Work done at Brown was supported through NSF DMR 0606264. Ò.I. acknowledges funding by the Spanish MINECO (projects MAT2009-0866 and MAT2012-33037), Catalan DURSI (project 2009SGR856), European Union FEDER funds (Una manera de hacer Europa) and CESCA and CEPBA under coordination of C4 for computer facilities.

REFERENCES

1.      Cortie, M. B.; McDonagh, A. M. Synthesis and Optical Properties of Hybrid and Alloy Plasmonic Nanoparticles. *Chemical Reviews* 2011, 111, 3713-3735.
2.      Armelles, G.; Cebollada, A.; Garcia-Martin, A.; Garcia-Martin, J. M.; Gonzalez, M. U.; Gonzalez-Diaz, J. B.; Ferreiro-Vila, E.; Torrado, J. F. Magnetoplasmonic nanostructures: systems supporting both plasmonic and magnetic properties. *Journal of Optics a-Pure and Applied Optics* 2009, 11.
3.      Leung, K. C.-F.; Xuan, S.; Zhu, X.; Wang, D.; Chak, C.-P.; Lee, S.-F.; Ho, W. K. W.; Chung, B. C. T. Gold and iron oxide hybrid nanocomposite materials. *Chemical Society Reviews* 2012, 41, 1911-1928.
4.      Pineider, F.; Fernandez, C. d. J.; Videtta, V.; Carlino, E.; al Hourani, A.; Wilhelm, F.; Rogalev, A.; Cozzoli, P. D.; Ghigna, P.; Sangregorio, C. Spin-Polarization Transfer in Colloidal Magnetic-Plasmonic Au/Iron Oxide Hetero-nanocrystals. *Acs Nano* 2013, 7, 857-866.
5.      Yu, H.; Chen, M.; Rice, P. M.; Wang, S. X.; White, R. L.; Sun, S. H. Dumbbell-like bifunctional Au-Fe3O4 nanoparticles. *Nano Letters* 2005, 5, 379-382.
6.      Wang, Z. F.; Shen, B.; Zou, A. H.; He, N. Y. Synthesis of Pd/Fe3O4 nanoparticle-based catalyst for the cross-coupling of acrylic acid with iodobenzene. *Chemical Engineering Journal* 2005, 113, 27-34.
7.      Xie, J.; Zhang, F.; Aronova, M.; Zhu, L.; Lin, X.; Quan, Q.; Liu, G.; Zhang, G.; Choi, K.-Y.; Kim, K.; Sun, X.; Lee, S.; Sun, S.; Leapman, R.; Chen, X. Manipulating the Power of an Additional Phase: A Flower-like Au-Fe3O4 Optical Nanosensor for Imaging Protease Expressions In vivo. *Acs Nano* 2011, 5, 3043-3051.
8.      Frey, N. A.; Srinath, S.; Srikanth, H.; Wang, C.; Sun, S. Static and dynamic magnetic properties of composite Au-Fe3O4 nanoparticles. *Ieee Transactions on Magnetics* 2007, 43.
9.      Lee, Y.; Angel Garcia, M.; Huls, N. A. F.; Sun, S. Synthetic Tuning of the Catalytic Properties of Au-Fe3O4 Nanoparticles. *Angewandte Chemie-International Edition* 2010, 49, 1271-1274.
10.     Frey, N. A.; Phan, M. H.; Srikanth, H.; Srinath, S.; Wang, C.; Sun, S. Interparticle interactions in coupled Au-Fe3O4 nanoparticles. *Journal of Applied Physics* 2009, 105.
11.     Iglesias, O.; Labarta, A.; Batlle, X. Exchange bias phenomenology and models of core/shell nanoparticles. *Journal of Nanoscience and Nanotechnology* 2008, 8, 2761-2780.
12.     Nogues, J.; Schuller, I. K. Exchange bias. *Journal of Magnetism and Magnetic Materials* 1999, 192, 203-232.
13.     Chandra, S.; Khurshid, H.; Phan, M. H.; Srikanth, H. Asymmetric hysteresis loops and its dependence on magentic anisotropy in exchange biased Co/CoO core-shell nanoparticles. *Appl. Phys. Lett.* 2012, 101.
14.     Zheng, R. K.; Liu, H.; Wang, Y.; Zhang, X. X. Cr2O3 surface layer and exchange bias in an acicular CrO2 particle. *Applied Physics Letters* 2004, 84, 702-704.
15.     Khurshid, H.; Chandra, S.; Li, W.; Phan, M. H.; Hadjipanayis, G. C.; Mukherjee, P.; Srikanth, H. Synthesis and magnetic properties of core/shell FeO/Fe3O4 nano-octopods. *Journal of Applied Physics* 2013, 113.
16.     Chandra, S.; Khurshid, H.; Li, W.; Hadjipanayis, G. C.; Phan, M. H.; Srikanth, H. Spin dynamics and criteria for onset of exchange bias in superspin glass Fe/gamma-Fe2O3 core-shell nanoparticles. *Physical Review B* 2012, 86.

17. Kodama, R. H.; Berkowitz, A. E.; McNiff, E. J.; Foner, S. Surface spin disorder in NiFe2O4 nanoparticles. *Physical Review Letters* 1996, 77, 394-397.
18. Peddis, D.; Cannas, C.; Piccaluga, G.; Agostinelli, E.; Fiorani, D. Spin-glass-like freezing and enhanced magnetization in ultra-small CoFe2O4 nanoparticles. *Nanotechnology* 2010, 21.
19. Mazo-Zuluaga, J.; Restrepo, J.; Munoz, F.; Mejia-Lopez, J. Surface anisotropy, hysteretic, and magnetic properties of magnetite nanoparticles: A simulation study. *Journal of Applied Physics* 2009, 105.
20. Lima, E.; Brandl, A. L.; Arelaro, A. D.; Goya, G. F. Spin disorder and magnetic anisotropy in Fe3O4 nanoparticles. *Journal of Applied Physics* 2006, 99.
21. Pal, S.; Chandra, S.; Phan, M.-H.; Mukherjee, P.; Srikanth, H. Carbon nanostraws: nanotubes filled with superparamagnetic nanoparticles. *Nanotechnology* 2009, 20.
22. Poddar, P.; Morales, M. B.; Frey, N. A.; Morrison, S. A.; Carpenter, E. E.; Srikanth, H. Transverse susceptibility study of the effect of varying dipolar interactions on anisotropy peaks in a three-dimensional assembly of soft ferrite nanoparticles. *Journal of Applied Physics* 2008, 104, 063901.
23. Bodker, F.; Morup, S.; Linderoth, S. Surface effects in metallic Iron nanoparticles. *Physical Review Letters* 1994, 72, 282-285.
24. Yang, T. Z.; Shen, C. M.; Li, Z.; Zhang, H. R.; Xiao, C. W.; Chen, S. T.; Xu, Z. C.; Shi, D. X.; Li, J. Q.; Gao, H. J. Highly ordered self-assembly with large area of Fe3O4 nanoparticles and the magnetic properties. *Journal of Physical Chemistry B* 2005, 109, 23233-23236.
25. Cabot, A.; Alivisatos, A. P.; Puntes, V. F.; Balcells, L.; Iglesias, O.; Labarta, A. Magnetic domains and surface effects in hollow maghemite nanoparticles. *Physical Review B* 2009, 79, 094419.
26. Khurshid, H.; Li, W.; Phan, M.-H.; Mukherjee, P.; Hadjipanayis, G. C.; Srikanth, H. Surface spin disorder and exchange-bias in hollow magnemite nanoparticles. *Appl. Phys. Lett.* 2012, 101, 022403.
27. Iglesias, O.; Batlle, X.; Labarta, A. Exchange bias and asymmetric hysteresis loops from a microscopic model of core/shell nanoparticles. *Journal of Magnetism and Magnetic Materials* 2007, 316, 140-142.
28. Poddar, P.; Wilson, J. L.; Srikanth, H.; Farrell, D. F.; Majetich, S. A. In-plane and out-of-plane transverse susceptibility in close-packed arrays of monodisperse Fe nanoparticles. *Physical Review B* 2003, 68.
29. Srikanth, H.; Wiggins, J.; Rees, H. Radio-frequency impedance measurements using a tunnel-diode oscillator technique. *Review of Scientific Instruments* 1999, 70, 3097-3101.
30. Aharoni, A.; Frei, E. H.; Shtrikman, S.; Treves, D. Bull.Res.Counc.Isr.,Sect.: 1957; Vol. F 6A, p 215.
31. Chandra, S.; Figueroa, A. I.; Ghosh, B.; Phan, M. H.; Srikanth, H.; Raychaudhuri, A. K. Phase coexistence and magnetic anisotropy in polycrystalline and nanocrystalline LaMnO3+delta. *Journal of Applied Physics* 2011, 109.
32. Wang, C.; Wei, Y.; Jiang, H.; Sun, S. Tug-of-War in Nanoparticles: Competitive Growth of Au on Au-Fe3O4 Nanoparticles. *Nano Letters* 2009, 9, 4544-4547.
33. Wang, H.; Zhu, T.; Zhao, K.; Wang, W. N.; Wang, C. S.; Wang, Y. J.; Zhan, W. S. Surface spin glass and exchange bias in Fe3O4 nanoparticles compacted under high pressure. *Physical Review B* 2004, 70.

34. Kobayashi, S.; Takahashi, S.; Kamada, Y.; Kikuchi, H. Strain-Induced Exchange Bias Effects in Chemically Ordered Pt3Fe Single Crystal. *Ieee Transactions on Magnetics* 2008, 44, 4225-4228.

35. Zhang, X. S.; Zhan, Q. F.; Dai, G. H.; Liu, Y. W.; Zuo, Z. H.; Yang, H. L.; Chen, B.; Li, R. W. Effect of mechanical strain on magnetic properties of flexible exchange biased FeGa/IrMn heterostructures. *Applied Physics Letters* 2013, 102.

36. Shendruk, T. N.; Desautels, R. D.; Southern, B. W.; van Lierop, J. The effect of surface spin disorder on the magnetism of gamma-$Fe_2O_3$ nanoparticle dispersions. *Nanotechnology* 2007, 18.

37. Iglesias, O.; Labarta, A. Finite-size and surface effects in maghemite nanoparticles: Monte Carlo simulations. *Physical Review B* 2001, 63, 184416.